\documentclass[twocolumn]{revtex4-1}%

\usepackage{amsfonts}
\usepackage{amsmath}
\usepackage{amssymb}
\usepackage{graphicx}%
\providecommand{\U}[1]{\protect\rule{.1in}{.1in}}
\providecommand{\U}[1]{\protect\rule{.1in}{.1in}}
\providecommand{\U}[1]{\protect\rule{.1in}{.1in}}

\begin{document}

\title{A Precise Measurement of the Oxygen Isotope Effect on the N\'{e}el temperature
in Cuprates}
\author{E.~Amit}
\email{Electronic address: eranamita@gmail.com}
\affiliation{Physics Department, Technion, Israel Institute of Technology, Haifa 32000, Israel.}
\author{A.~Keren}
\affiliation{Physics Department, Technion, Israel Institute of Technology, Haifa 32000, Israel.}
\author{J.~S.~Lord}
\affiliation{Rutherford Appleton Laboratory, Chilton Didcot, Oxfordshire OX11 0QX, U.K.}
\author{P.~King}
\affiliation{Rutherford Appleton Laboratory, Chilton Didcot, Oxfordshire OX11 0QX, U.K.}

\begin{abstract}
A limiting factor in the ability to interpret isotope effect measurements in cuprates is the absence of
sufficiently accurate data for the whole phase diagram; there is precise data for $T_{c}$, but not for
the antiferromagnetic transition temperature $T_{N}$. Extreme sensitivity of $T_{N}$ to small
changes in the amount of oxygen in the sample is the major problem. This problem is solved here by
using the novel compound (Ca$_{0.1}$La$_{0.9}$)(Ba$_{1.65}$La$_{0.35}$)Cu$_{3}$O$_{y}$ for which there is a region where $T_{N}$ is independent of oxygen doping.
Meticulous measurements of $T_{N}$ for samples with $^{16}$O and
$^{18}$O find the absence of an oxygen isotope effect on $T_{N}$ with unprecedented accuracy. A possible interpretation of our finding and existing data is that isotope substitution affects the normal state charge carrier density.
\end{abstract}
\maketitle

Isotope substitution is a powerful experimental tool used
to investigate complex systems. Ideally, the isotope substitution affects only
one parameter, for example, a phonon frequency which is directly related to
the nuclear mass. However, the strong coupling of many parameters in cuprates highly limit the ability to interpret isotope effect (IE) experiments.
Eventhough the oxygen isotope substitution is known to affect the superconductivity transition temperatures $T_{c}$, it is unclear whether it primarily impacts
phonons, polarons, magnons, doping or other physical properties.\cite{Lee}

The isotope effect is usually described using the isotope coefficient
$\alpha$ via the relation
\begin{equation}
{T_{q}\propto M^{-\alpha_{q}}} \label{alpha}%
\end{equation}
where $T_{q}$ is a phase transition temperature, $M$ is the isotope mass and $q=C, N,$ and $g$ for the superconducting (SC), antiferromagnetic (AFM) and spin glass critical temperatures, respectively. In many conventional superconductors $\alpha$ was found to be very close to
0.5.\cite{Maxwell, Reynolds} The explanation of this $\alpha$ in terms of
Cooper pairs glued by phonons was one of the triumphs of the BCS theory for
metallic superconductors.\cite{BCS}

In cuprates the isotope effect is much more complicated and $\alpha$ is not
single valued and varies across the phase diagram. The consensus today for YBCO like compounds is that in the SC phase, close to optimal doping, the oxygen IE is
\begin{equation}
\alpha_{C}^{od}=0.018\pm0.005, \label{alphaOD}%
\end{equation}
which is very small but non zero.\cite{Khasanov} On the SC dome
$\alpha_{C}$ increases as the doping decreases.\cite{Pringle, Keller} In
the glassy state less data is available, but it seems that the isotope
effect reverses sign and $\alpha_{g}$ becomes negative. In extremely
underdoped samples, where long range AFM order prevails at
low temperatures, data is scarce, controversial, and has relatively large
error bars.\cite{Khasanov, Zhao} The most recent measurements with Y$_{y}%
$Pr$_{1-y}$Ba$_{2}$Cu$_{3}$O$_{7-\delta}$ show that $\alpha_{N}=0.02(3)$ in
the parent compound.\cite{Khasanov} There are also several theories dealing
with the variation of $\alpha$ along the phase diagram,\cite{Zhao2, A. Bill,
J. F. Baugher, D.Fisher, Kresin, Serbyn} but since $\alpha_{C}^{od}$ and
$\alpha_{N}$ are within an error bar of each other one cannot contrast these
theories with experiments. In particular, it is impossible to tell whether the
same glue that holds the spins together holds the cooper pair together, or
not. Increasing the accuracy of the IE measurements of the N\'{e}el temperature will shed
light on the role of isotope substitution.%

\begin{figure}
\begin{center}
\includegraphics[width=8cm]{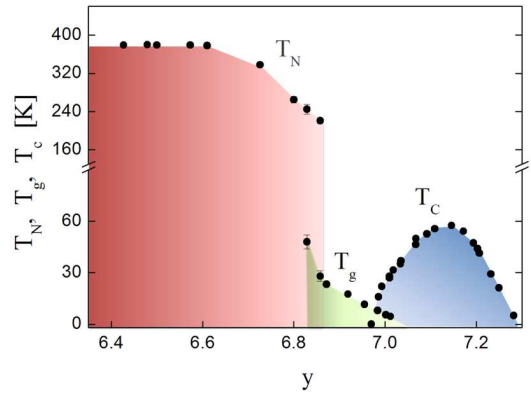}%
\caption{The (Ca$_{0.1}$La$_{0.9}$)(Ba$_{1.65}$La$_{0.35}%
$)Cu$_{3}$O$_{y}$ phase diagram \cite{ROfer}. The
antiferromagnetic, spin glass, and superconducting phases are represented in
red, green and blue, respectively. For $y<6.6$ the N\'{e}el temperature
does not depend on $y$. This enables accurate measurements of the oxygen
isotope effect on the N\'{e}el temperature $T_{N}$.}%
\label{x01}%
\end{center}
\end{figure}

As mentioned above, a major limitation on the measurements of $\alpha_{N}$ is
the strong dependence of $T_{N}$ on doping. For example, in Y$_{y}$Pr$_{1-y}%
$Ba$_{2}$Cu$_{3}$O$_{7-\delta}$, $T_{N}$ decreases with increasing doping at a
rate $\Delta T_{N}=2.5$~K per $\Delta y=0.01$. This strong temperature
dependence is, of course, common to many other cuprates. As a consequence, it
is very difficult to prepare two samples with exactly the same $T_{N}$ even
with the same isotope. The smallest fluctuation in either $y$ or $\delta$ may
lead to a huge fluctuation in $T_{N}$ regardless of the isotope effect. This
is not the case for the (Ca$_{0.1}$La$_{0.9}$)(Ba$_{1.65}$La$_{0.35}$)Cu$_{3}%
$O$_{y}$ where $T_{N}$ is constant for oxygen density $y<6.6$, as can be seen in Fig.~\ref{x01}.\cite{ROfer}

For our experiments, four sintered pellets were prepared using standard
techniques.\cite{Goldschmidt} Two of the pellets were enriched with $^{18}$O
isotope and two with $^{16}$O isotope in the same procedure: The samples were
placed simultaneously in two closed tubes, each with different isotope gas, and then they were heated to allow the isotope to diffuse into the sample. In order to
achieve a higher percentage of gas, the enrichment was repeated several times.

\begin{figure}
\begin{center}
\includegraphics[width=8cm]{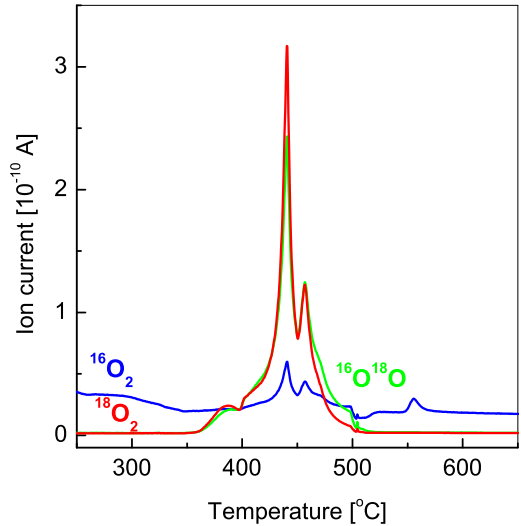}
\end{center}
\caption{Ion current signals for the $^{18}$O$_{2}$, $^{18}$O$^{16}$O and
$^{16}$O$_{2}$ molecules (red, green and blue color, respectively) obtained
during heating of the sample in a stream of helium. This graph shows that the
$^{18}$O isotope fraction in the samples is bigger than 70\% (see text).}%
\label{MS}%
\end{figure}

The $^{18}$O isotope content in the samples was determined based on
measurements of gas composition being in equilibrium with the sample during
the exchange. Balzers Prisma mass spectrometer was used to analyze in situ
isotopic composition of the atmosphere. After the exchange process was
performed, the weight increase of the sample was also determined as the light
$^{16}$O isotope was exchanged with the heavy $^{18}$O. The isotope enrichment
in the samples measured by both methods looked to be higher than 80\%. Finally
a Thermal Analysis experiment (TA) was performed for the investigated samples
after all experiments described in this work were fulfilled in order to verify
the isotope enrichment. The samples were heated up to 1200$^{\circ}$C in the
NETSCH STA 449C Jupiter analyzer in a stream of helium. During the TA
experiments, ion current signals for the $^{18}$O$_{2}$, $^{18}$O$^{16}$O and
$^{16}$O$_{2}$ molecules were measured using a mass spectrometer (ThermoStar
Pfeiffer Vacuum). The results are shown on Fig.~\ref{MS}. The isotope content
deduced from these measurements (comparing peak areas for the signals of
particular oxygen molecules) was larger than 70\%.

\begin{figure}
\begin{center}
\includegraphics[
natheight=4.804900in,
natwidth=3.521500in,
height=4.3855in,
width=3.2214in
]%
{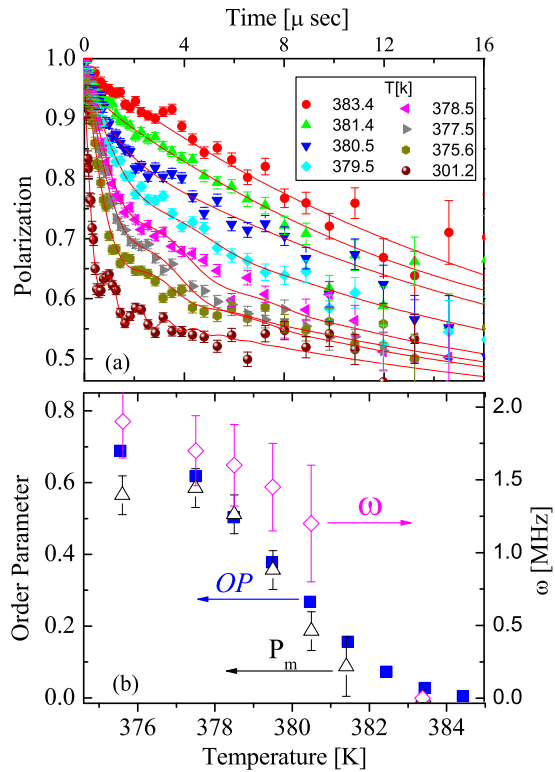}%
\caption{(a) The $\mu$SR raw data at different temperatures and
fits of Eq.~\ref{muonfit} to the data in solid lines. (b) Comparison between
three different methods used to describe the AFM phase transition: (I) the
muon precession frequency $\omega$ obtained from Eq.~\ref{muonfit}. (II) the
magnetic volume fraction $P_{m}$ determined by Eq.~\ref{muonfit}, and (III)
$\mathcal{OP}$ calculated by Eq.~\ref{muonopuse}.}%
\label{MuSR raw}%
\end{center}
\end{figure}

The oxygen IE was measured using zero-field muon spin rotation/relaxation
($\mu$SR). We particularly used the ISIS facility, which allows low muon
relaxation rate and rotation frequency measurements. This is ideal for
measurements near magnetic phase transitions where the muon signal varies on a
long time scale. $\mu$SR data at temperatures close to the phase transition
are shown in Fig.~\ref{MuSR raw}. As the temperature is lowered from 383.4~K the
relaxation rate increases. At $T=$378.5~K oscillations appear in the data
indicating the presence of long range magnetic order. The frequency of
oscillations and the relaxation rate increase as the temperature is further
lowered. The formula
\begin{equation}
P_{z}(t)=P_{m}(ae^{-\lambda_{1}t}+(1-a)e^{-\lambda_{2}t}cos(\omega
t))+P_{n}e^{-\Delta t} \label{muonfit}%
\end{equation}
was fitted to the muon polarization, where $P_{m}$, $\lambda_{1}$,
$\lambda_{2}$ and $\omega$ are the polarization, relaxation rates, and
frequency of muons spin in the fraction of the samples which is magnetic, and
$a$ is the weighting factor between the muons experiencing transverse field
and longitudinal field. This factor is close to 2/3 and
temperature-independent. $P_{n}$ and $\Delta$ are the polarization and
relaxation of the spin of muons that stopped in the non-magnetic volume of the
sample. The solid lines in Fig.~\ref{MuSR raw}(a) represent the fits.%

\begin{figure}
\begin{center}
\includegraphics[
natheight=3.409100in,
natwidth=4.455500in,
height=2.6455in,
width=3.4523in
]%
{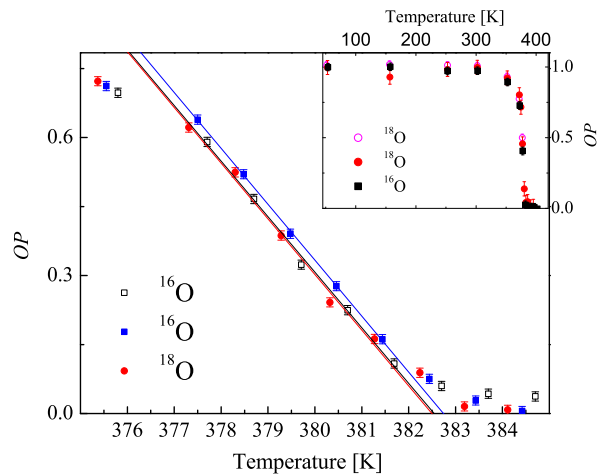}%
\caption{The parameter $\mathcal{OP}$ (see Eq.~\ref{muonopuse})
versus temperature near $T_{N}$ for $^{18}$O and $^{16}$O rich samples. The
solid lines are fits to the data near the phase transision used to determine
$T_{N}$. The inset shows a second experiment on the entire
temperature range. No oxygen isotope effect of $T_{N}$ is observe within
experimental error. }%
\label{MuSR OIE}%
\end{center}
\end{figure}

The AFM order parameter is determined by the frequency $\omega=\gamma B$,
where $B$ is the average magnetic field at muon site and $\gamma$ is the muon
gyromagnetic ratio. This frequency can be easily extracted from the $\mu$SR
data well below the transition but is very difficult to determined near the
transition. A second approach is to treat the magnetic volume fraction $P_{m}$
as the order parameter.\cite{ROfer2,Khasanov} $P_{m}$ can be followed more
closely to $T_{N}$, although this parameter also has large error bars when
$\omega$ is not well defined. We used an alternative approach similar to Ref.
\cite{Shay}; we define an order parameter which, does not require a fit, via the
relation
\begin{equation}
\mathcal{OP}(T)\equiv\frac{<P_{inf}>-<P(T)>}{<P_{inf}>-<P(0)>}
\label{muonopuse}%
\end{equation}
where $<P(T)>$ is the average polarization at temperature $T$, and $<P_{inf}>$
is the average polarization above the transition. The denominator normalizes
$\mathcal{OP}$ to 1 at zero temperature. All three order parameters for one
sample are shown in Fig.~\ref{MuSR raw}(b). The transition temperatures
determined using the different order parameters are in good agreement. The
advantages of $\mathcal{OP}$ are clear: it is a model free and has very small uncertainties.%

In Fig.~\ref{MuSR OIE} we present the $\mathcal{OP}$ for two samples with
$^{16}$O and one with $^{18}$O around 380~K. A wide temperature range from 50
to 410~K is shown in the inset for an experiment done on separate occasion, in
which two samples of $^{18}$O and one with $^{16}$O were examined. We
determine $T_{N}$ by fitting a straight line to the data in the main panel of
Fig.~\ref{MuSR OIE} in the temperature range 378 to 382~K, for each sample,
and taking the crossing with the temperature axis. We find that $T_{N}^{18}%
$=382.49(0.34)~K and $T_{N}^{16}$=382.64(0.29)~K. For 100\% isotope
substitution the isotope exponent is determined by
\begin{equation}
\alpha_{N}=-\frac{M_{o}^{16}}{T_{N}^{16}}\frac{T_{N}^{18}-T_{N}^{16}}%
{M_{o}^{18}-M_{o}^{16}}. \label{alphaDef}%
\end{equation}
When taking into account the isotopic fraction in the samples we obtain:
\begin{equation}
\alpha_{N}=0.005\pm0.011. \label{alphaN}%
\end{equation}
This result indicates that $\alpha_{N}<\alpha_{C}^{od}$ (see Eq.~\ref{alphaOD}%
) beyond the error bars, and is consistent with no isotope effect on $T_{N}$.%

\begin{figure}
\begin{center}
\includegraphics[
height=4.0318in,
width=2.7415in
]%
{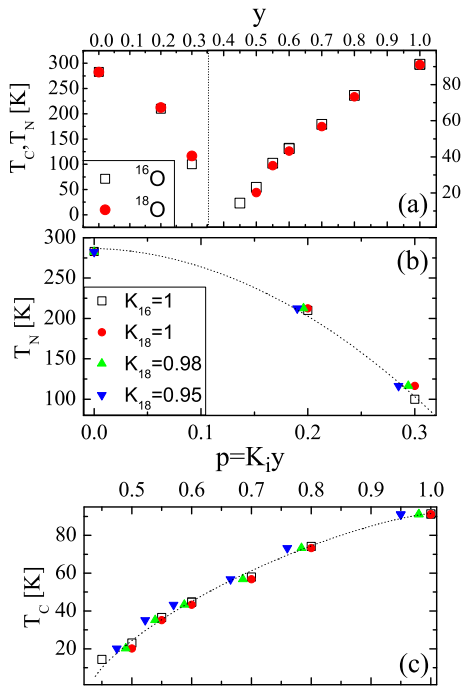}%
\caption{(a): Oxygen isotope effect of T$_{N}$ (left) and
T$_{c}$ (right) in Y$_{y}$Pr$_{1-y}$Ba$_{2}$Cu$_{3}$O$_{7-\delta}$ taken from
Ref.~\ref{Khasanov phase diagram}. The $^{18}$O and $^{16}$O samples are in
filled red circles and empty black squares, respectively. Figure (b) and (c)
are demonstration of the IE using charge carriers density $p=K_{i}y$, insted
of oxygen content $y$. The values of $K_{i}$ \ are given in the the figures.
Green triangles ($K_{i}=0.98$) represent reduction of 2\% in the number of
charge carriers in the $^{18}$O samples compared to the $^{16}$O sample. In
this case both $T_{N}$ and $T_{c}$ are functions of $p$ regardless of the
isotope.}%
\label{Khasanov phase diagram}%
\end{center}
\end{figure}

One possible interpretation of these results is that magnetic excitations are
not relevant for superconductivity since the isotopes affect $T_{c}$
without affecting $T_{N}$. This approach was presented, for example, by Zhao
\emph{et al.}.\cite{Zhao2} They found that samples enriched with $^{18}$O have longer penetration depth $\lambda$ than samples enriched with $^{16}$O with the same amount of oxygen per unit cell. $\lambda$ is related to
the SC carrier density $n_{s}$ and effective mass $m^{\ast}$ by
$\lambda^{-2}\propto n_{s}/m^{\ast}$ so a priory both $n_{s}$ and $m^{\ast}$
can be affected by isotope substitution.\cite{Uemura} They ruled out the possibility that the number of carriers concentration varies by demonstrating that the thermal
expansion coefficient of samples with different isotopes are the same. The
authors therefore concluded that the IE changes the mass of the cooper pairs,
which could be explained by polaronic supercarriers.

An alternative interpretation is that the isotopes affect the efficiency of
doping, as suggested in Ref.~\cite{Kresin}. To demonstrate this interpretation we
present in Fig.~\ref{Khasanov phase diagram}(a) the critical temperatures in
Y$_{y}$Pr$_{1-y}$Ba$_{2}$Cu$_{3}$O$_{7-\delta}$ for the two different isotopes
taken from Ref.~\cite{Khasanov}. $T_{N}^{18}$ seems to be a bit higher than
$T_{N}^{16}$, but $T_{c}^{18}$ is a bit lower than $T_{c}^{16}$. However, if
we define an efficiency parameter $K_{i}$ which relates the number of holes
$p$ to the number of oxygens in the unit cell $y$ via $p=K_{i}y$, where $i$
stands for the isotope type, we can generate a unified phase diagram. This is
demonstrated in Fig.~\ref{Khasanov phase diagram}(b) and
\ref{Khasanov phase diagram}(c) for $T_{N}$ and $T_{c}$, respectively. In
these graphs three different values of $K_{18}$ are used to generate $p$ while
keeping $K_{16}=1$. When using $K_{18}=0.98$, both curves of $T_{N}$ and
$T_{c}$ versus $p$ for the two different isotopes collapse to the same curve. Similar scaling of the doping axis was applied to the (Ca$_{x}$La$_{1-x}$)(Ba$_{1.75-x}$La$_{0.25+x}$)Cu$_{3}$O$_{y}$ system
\cite{ROfer} and was explained by NQR.\cite{Amit}

Next we discuss the IE on the stiffness in the above scenario. We assume that the effective mass of the SC charge carrier $m^{\ast}$, the critical $p$ where superconductivity starts $p_{crit}$, and where $T_c$ is optimal $p_{opt}$, are not affected by the isotope substitution as suggested by Fig.~\ref{Khasanov phase diagram}(c) and demonstrated in Ref.~\cite{Amit}. The stiffness can be measured by the muon transverse relaxation rate $\sigma$, and it is expected that:\cite{Uemura}
\begin{equation}
\sigma^i=C(p^i-p_{crit}) \label{sigma1}%
\end{equation}
where $C$ is a constant. Dividing the differential of $\sigma$ from Eq.~\ref{sigma1} by the relaxation at optimal doping yields:
\begin{equation}
\frac{d\sigma}{\sigma_{opt}}=\frac{dp}{p_{opt}-p_{crit}}=\frac{y(K_{16}-K_{18})}{p_{opt}-p_{crit}} \label{sigma2}
\end{equation}

The expected change in the stiffness due to isotope substitution can be calculated from Eq.~\ref{sigma2} using $p_{opt}=1$, $p_{crit}=0.42$ (which are extracted from Fig.~\ref{Khasanov phase diagram}(b)) and $\sigma_{opt}=3.0$~$\mu s^{-1}$.\cite{Khasanov2} For $y=0.8$, $K_{16}=1$ and $K_{18}=0.98$ we get $d\sigma=0.083$~$\mu s^{-1}$. This value is consistent with the experimental value of $d\sigma=0.08(1)$~$\mu s^{-1}$ reported in Ref.~\cite{Khasanov2}. In other words a 2\% difference in the doping efficiency between the two isotopes can explain both the variations in phase diagram in Fig.~\ref{Khasanov phase diagram}(a) and the variation in the stiffness.

Our experiment shows that oxygen isotope substitution does not
affect the N\'{e}el temperature and therefore does not play a role in magnetic
excitations. However, the isotope effect of $T_{c}$ does not necessarily imply
that phonons play a role in cuprate superconductivity. We show that an isotope
dependent doping efficiency can explain the variation in $T_{c}$ and in the magnetic penetration depth between
samples rich in $^{16}$O or $^{18}$O.
We would like to thank the ISIS pulsed muon facility at Rutherford Appleton
Laboratory, UK for excellent muon beam conditions. This work was funded by the
Israeli Science Foundation and the Posnansky research fund in high temperature superconductivity.

\end{document}